\documentclass[10pt,letterpaper]{article}
\usepackage{opex3}

\begin{document}
\title{Broadband time-reversal of optical pulses using a switchable photonic-crystal mirror}

\author{Yonatan Sivan and John B. Pendry}

\address{The Blackett Laboratory, Department of Physics, Imperial College London, London SW72AZ}

\email{ysivan@imperial.ac.uk} %% email address is required

\begin{abstract}
Recently, Chumak {\em et al.} have demonstrated experimentally the time-reversal of microwave spin pulses based on non-adiabatically tuning the wave speed in a spatially-periodic manner~\cite{spin-wave-reversal-chumak}. Here, we solve the associated wave equations analytically, and give an explicit formula for the reversal efficiency. We then discuss the implementation for short optical electromagnetic pulses and show that the new scheme may lead to their accurate time-reversal with efficiency higher than before.
\end{abstract}


% \begin{verbatim}
% \begin{figure}[htbp]
% %\centering\includegraphics[width=7cm]{opexfig1}
% \caption{Sample caption (Ref. \cite{Oron03}, Fig. 2).}
% \end{figure}

% \begin{equation}
% H = \frac{1}{2m}(p_x^2 + p_y^2) + \frac{1}{2} M{\Omega}^2
%      (x^2 + y^2) + \omega (x p_y - y p_x).
% \end{equation}
% \end{verbatim}

Time-reversal is one of the most spectacular and useful wave phenomena. A time-reversed pulse evolves as if time runs backward, thus eliminating any distortions or scattering that occurred at earlier times. This enables light detection, imaging and focusing through complex media~\cite{Fink_review,Mosk-Katz} with
applications in diverse fields such as medical ultrasound~\cite{Fink_review}, communication systems and adaptive optics~\cite{Pepper-book}, superlensing~\cite{tr_super_lens_pendry}, ultrafast plasmonics~\cite{Stockman_reversal}, biological and THz imaging~\cite{yaqoob_Nat_Phot}, spintronics~\cite{magnonic_crystals_transmission_lines} and quantum information and computing~\cite{Cucchietti}.

For low frequency waves (e.g., in acoustics, microwave spin/electromagnetic waves etc.), time-reversal can be accomplished by electronic sampling, recording, and playing back~\cite{Fink_review}. This is possible since in this frequency range, the pulse oscillates on a scale slower than electronic sampling speed. On the other hand, for high frequency waves, specifically, for optical electromagnetic waves, the pulse oscillations (and even the pulse envelope itself for sufficiently short pulses) are too short to be sampled properly by even the fastest electronic detector. % Moreover, for sufficiently short pulses, even the pulse's envelope is too short to be sampled electronically.

The standard solution in the optical regime is to use nonlinear processes such as Three-Wave or Four-Wave Mixing (3WM or 4WM, respectively), see e.g.,~\cite{Pepper-book,Weiner_holography,Marom_Fainman,Kuzucu_Gaeta}. However, while such techniques have been demonstrated experimentally, they usually suffer from one or several disadvantages. In particular, they require fairly high intensities, thus, limiting on-chip integration; almost all existing schemes are narrowband % , hence, their applicability to ultrashort pulses is limited; finally, 
whereas the schemes which are applicable to relatively short pulses are in general complex, requiring complicated setups and sometimes even cryogenic temperatures. Finally, some schemes may be difficult to apply to~2- and 3-dimensional systems.

Recent suggestions to overcome some of these limitations involved dynamically-tuned optical periodic structures, known as photonic crystals (PhCs)~\cite{Fan-reversal,Longhi-reversal}. However, despite being very efficient, the structures and/or modulations required for these schemes were quite challenging to realize; in addition, these schemes allow for reversal of only relatively narrow pulses. However, these obstacles were recently lifted in a study by Chumak {\em et al.}~\cite{spin-wave-reversal-chumak}. In particular, it was shown in~\cite{spin-wave-reversal-chumak} that a pulse propagating in a homogeneous material can be time-reversed if the material is subject to a spatially-periodic, time-localized {\em non-adiabatic} modulation. This scheme is closely related to other non-adiabatic optical pumping based schemes which were studied theoretically earlier in slab waveguides~\cite{Miller_OL,Tsang_Psaltis2}, atomic vapours~\cite{Matsko} % what is the efficiency? what is the spectral bandwidth? probably small. % and also in spin waves - what is the efficiency? At the moment i avoid Busch - complicated 2D PhC + unclear efficiency.
and zero-gap photonic-crystals~\cite{Sivan-Pendry-letter,Sivan-Pendry-article}.

% CURRENTLY I IGNORE THE FACT THEY PERFORM WAVE FRONT REVERSAL...

% This was experimentally demonstrated for microwave spin waves of a relatively narrow spectral width. 

% a detailed discussion on efficiencies, dependence on relative durations of pump and signal pulses and a comparison to previous schemes was not provided.

In this article, we study the time-reversal scheme of~\cite{spin-wave-reversal-chumak} theoretically and compare it to previous non-adiabatic modulation-based schemes; we focus on time-reversal of ultrashort optical pulses, which as noted above, still has not been demonstrated experimentally. We first give a simple heuristic explanation of the dynamically-tuned PhC-based schemes~\cite{spin-wave-reversal-chumak,Sivan-Pendry-letter,Sivan-Pendry-article} showing that the reversal originates from switching-on a bulk PhC mirror (Section~\ref{sec:principles}). Then, we provide a detailed analytical study of the reversal process in~\cite{spin-wave-reversal-chumak}, derive a simple explicit formula for the reversal efficiency in the weak-coupling limit and compare it with existing techniques (Section~\ref{sec:analysis}). We then briefly discuss the implementation for optical electromagnetic pulses (Section~\ref{sec:implementation}). % with emphasis on efficiency and spectral width of the reversed pulses.

\section{Principles of time-reversal using a switchable mirror}\label{sec:principles}
% In this study, we offer an alternative approach which overcomes all those limitations. % Our scheme is based on switchable photonic-crystal mirrors.
In order to understand the reversal schemes of~\cite{spin-wave-reversal-chumak,Sivan-Pendry-letter,Sivan-Pendry-article}, it is beneficial to adopt a somewhat heuristic interpretation. Recall that when a pulse is reflected by a (standard) mirror, its spatial components change their propagation direction at different times, i.e., the leading edge first and trailing edge last. Thus, the pulse effectively undergoes a {\em U-turn}, i.e., the leading edge remains the leading edge etc.. Now, imagine that one could change of direction of the pulse propagation {\em at all points in space and at the same time}. Then, obviously, the leading edge will become the trailing edge, and vice versa, i.e., the pulse is (time-) reversed. % This scenario corresponds to time-reversal of the pulse.

In order to perform such an extreme manipulation, one needs to abruptly reduce the transmissivity of the medium, preferably to zero, everywhere is space and for {\em a spectral band as wide as possible}. Possibly the simplest way to do that would be to open a frequency bandgap by periodically modulating the material properties. % dynamically modulate the material properties in a periodic manner. Indeed, materials with periodically varying properties give rise to bandgaps which are energy/frequency regimes in which wave propagation is forbidden.
Heuristically, when the bandgap is turned on, the wave cannot propagate in any direction. Instead, the forward waves are then repeatedly converted to backward waves, then back to forward waves and vice versa. If one re-establishes the transmissivity once most of the energy of the forward wave has been converted to a backward wave, then a time-reversed pulse is released backwards. In a sense, this procedure transforms a perfectly transmitting medium into a ``volume'' mirror. Accordingly, in what follows we refer to these schems as switchable mirror (SM) -based reversal schemes.

This heuristic explanation clarifies why the zero-gap-based switchable-mirror (ZGSM) is equivalent to a homogeneous-medium-based switchable-mirror (HSM). Indeed, unlike a finite-gap system, in both structures all the incident light is perfectly admitted. In addition, in both structures a gap is opened due to the modulation (see~\cite{spin-wave-reversal-chumak,Sivan-Pendry-article}). It is also implied by the heuristic explanation (and later proved analytically in Section~\ref{sec:analysis}), that in both ZGSM and HSM, {\em although the wave-front is reversed, it is not conjugated}. Thus, the scheme can lead to perfect time-reversal only if it is complemented by a consequent step of phase-conjugation, e.g., via nearly-degenerate 4WM~\cite{Sivan-Pendry-letter,Sivan-Pendry-article}. As shown in Section~\ref{sec:analysis}, this may also be beneficial in order to make the scheme more efficient.

The advantage of the SM-based schemes is the ability to reverse pulses of almost unlimited wide spectrum. Indeed, the only limitation is the shortness of the modulation. Also, as noted in~\cite{spin-wave-reversal-chumak,Sivan-Pendry-letter,Sivan-Pendry-article}, the new reversal schemes, which require only a periodic modulation rather than complex optics-specific concepts, open the way for time-reversal in many other wave systems for which time-reversal was not accessible before, such as atomic physics~\cite{Matsko}, quantum computing~\cite{Cucchietti} etc.. 

Importantly, the HSM has several advantages over the ZGSM. First, the former is obviously simpler as essentially no fabrication is required; its performance is also practically insensitive to the modulation spatial pattern. Second, our analysis (Eq.~(\ref{eq:E-sol}) below) shows that the reversal effeciency of the in a HSM is about an order of magnitude higher than in a ZGSM. Third, while the frequency conversion performed in a ZGSM is purely vertical~\cite{Sivan-Pendry-letter,Sivan-Pendry-article}, the frequency conversion in a HSM involves also a change in the sign of the carrier wavevector, i.e., the dynamic grating provides the momentum to allow a horizontal transition between positive to negative momentum~\cite{spin-wave-reversal-chumak}. % Thus, the phase velocity of the reversed wave in the HSM is negative, but it is positive in the ZGSM.
As a result, the out-coupling of pulses from the HSM is not plagued by reflections at the boundaries, as may happen in a ZGSM. A final advantage of the HSM is that by choosing the proper modulation pattern, the reversal can be performed for any angle of incidence, for plane-waves as well as for beams, for high dimensional gratings or waveguide structures as well as for any incident carrier frequency. % Thus, overall, the HSM offers a very attractive route for time-reversal of short optical pulses.

% {\bf Add figures of transitions}

% The efficiencies can be up to an order of magnitude higher than those offered by previously suggested schemes. 

% very simple structures are required which do not require any phase-matching considerations, only a linear modulator or a single pump pulse are required etc..

\section{Analysis}\label{sec:analysis}
In this Section we derive the coupled envelope equations and solve them analytically in the weak-coupling limit. Since the former step is very similar to the derivations in~\cite{shallow_grating_review}, and latter step is almost identical to the procedure detailed in~\cite{Sivan-Pendry-letter,Sivan-Pendry-article}, here, the derivation and solution are given quite briefly.
 
Consider an electromagnetic plane-wave pulse propagating along the $x$-direction in a homogeneous medium which is time-modulated in the following manner
\begin{equation}\label{eq:eps}
\epsilon(x,t) = n_0^2 + \Delta \epsilon\ m(t - t_0) \epsilon_m(x).
\end{equation}
Here, $n_0$ is the average refractive index, the modulation is spatially periodic, i.e.,
\begin{equation}
\epsilon_m(x) = \epsilon_m(x + d), \quad \quad \int_0^d \epsilon_m(x) dx = 0, \quad \quad max[\epsilon_m(x)] = 1, \nonumber
\end{equation}
and time-localized around $t_0$ with $max[m(t - t_0)] = m(t_0) = 1$, i.e., the modulation essentially turns on a periodic grating perpendicular to the direction of propagation of the pulse. In this case, the Maxwell equations reduce to the 1D wave equation
\begin{eqnarray}\label{eq:waeq}
\partial_{xx} E(x,t) = \frac{1}{c^2} \partial_{tt} \left[\epsilon(x,t) E(x,t)\right].
\end{eqnarray}

In order to solve Eq.~(\ref{eq:waeq}), we follow the derivation given in~\cite{shallow_grating_review}, used previously to study the propagation of optical pulses in a photonic crystal with a cubic (Kerr) nonlinear response\footnote{Note that unlike our previous studies of time-reversal in a zero-gap photonic crystal system~\cite{Sivan-Pendry-letter,Sivan-Pendry-article}, which can only be studied within the deep-grating formalism~\cite{deep_gratings_PRE_96}, here, we can use the shallow-grating analysis~\cite{shallow_grating_review}. The latter is much simpler to derive, and does not require the use of multiple scales analysis and calculation of the Floquet-Bloch modes which are required in~\cite{Sivan-Pendry-letter,Sivan-Pendry-article}. }. We assume the field can be written as % a sum of a forward and backward components, each given by the product of a carrier wave and a SVE envelope.
\begin{eqnarray}\label{eq:ansatz}
E(x,t) &=& \left[E^+(x,t) e^{i k_0 x} + E^-(x,t) e^{- i k_0 x}\right] e^{- i \omega_0 t} + c.c.,
\end{eqnarray}
where $E^\pm$ represent the Slowly-Varying Envelopes (SVEs) of the forward and backward field compoenets, respectively; $\omega_0 = c k_0$, $k_0 = 2 \pi n_0/\lambda_v$ and $n_0$ are the carrier frequency, wavevector and refractive index of the medium, respectively, with $\lambda_v$ being the vacuum wavelength. Substituting the ansatz~(\ref{eq:ansatz}) into Eq.~(\ref{eq:waeq}), % gives
%\begin{eqnarray}\label{eq:deriv1}
%E^+_{xx}(x,t) &+& 2 i k_0 E_x^+ - k_{c,x}^2 E^+ + \left(E^-_{xx} - 2 i k_0 E_x^- - k_{c,x}^2 E^-\right)e^{- 2 i k_0 x} \nonumber \\
%% &=& \frac{1}{c^2} \left[\epsilon E_{tt} + 2 \epsilon_t E_t + \epsilon_{tt} E\right] = \frac{n_0^2}{c^2} E_{tt} + M_0 \frac{\epsilon_m(x)}{c^2} \left[m E_{tt} + 2 m_t E_t + m_{tt} E\right] \nonumber \\
%&=& \frac{n_0^2}{c^2} \left(- \omega_0^2 E^+ - 2 i \omega_0 E^+_t + E^+_{tt}\right) + \frac{n_0^2}{c^2} \left( - \omega_0^2 E^- - 2 i \omega_0 E^-_t + E^-_{tt}\right)e^{- 2 i k_{c,x} x} \nonumber \\
%&+& \Delta \epsilon \frac{\epsilon_m(x)}{c^2} \left[m_{tt} E^+ + 2 m_t \left(E^+_t - i \omega_0 E^+\right) + m \left(- \omega_0^2 E^+ - 2 i \omega_0 E^+_t + E^+_{tt}\right)\right] \nonumber \\
%&+& \Delta \epsilon \frac{\epsilon_m(x)}{c^2} \left[m_{tt} E^- + 2 m_t \left(E^-_t - i \omega_0 E^-\right) + m \left(- \omega_0^2 E^- - 2 i \omega_0 E^-_t + E^-_{tt}\right)\right]e^{- 2 i k_0 x}. \nonumber \\
%\end{eqnarray}
neglecting the c.c. terms and the second order derivative terms, and removing the factor $e^{- i \omega_0 t}$ gives
\begin{eqnarray}\label{eq:deriv2}
2 i k_0 E_x^+ &+& 2 i \omega_0 \frac{n_0^2}{c^2} E^+_t = 2 i \left(k_0 E_x^- - \omega_0 \frac{n_0^2}{c^2} E^-_t\right)e^{- 2 i k_0 x} \nonumber \\
&+& \Delta \epsilon \frac{\epsilon_m(x)}{c^2} \left[2 m_t \left(E^+_t - i \omega_0 E^+\right) - m \left(2 i \omega_0 E^+_t + \omega_0^2 E^+\right)\right] \nonumber \\
&+& \Delta \epsilon \frac{\epsilon_m(x)}{c^2} \left[2 m_t \left(E^-_t - i \omega_0 E^-\right) - m \left(2 i \omega_0 E^-_t + \omega_0^2 E^-\right)\right]e^{- 2 i k_0 x}.
\end{eqnarray}
The spatial modulation $\epsilon_m(x)$ couples the forward and backward field components. When $k_0$ is close to the first bandgap, i.e., when $k_0 = k^{(g)} + \delta k$ (with $k^{(g)} \equiv \pi/d$ or equivalently, $\lambda_v^{(g)} = 2 n_0 d$), the $j = \pm 1$ components of the grating are close to the phase mismatch between the forward and backward field components, so that the coupling becomes most efficient. Following~\cite{shallow_grating_review}, we now expand the spatial part of the modulation as a Fourier series as follows % only relevant factor is the period of the structure
\begin{equation}\label{eq:eps_m}
\epsilon_m(x) = \sum_{j=1}^\infty \epsilon_m^{(j)} e^{i \frac{2 \pi j x}{d}} + c.c., \quad \quad \epsilon_m^{(j)} = \frac{1}{d} \int_0^d \epsilon_m(x)  e^{i \frac{2 \pi j x}{d}} dx. \nonumber
\end{equation}
For a weak grating, $\Delta \epsilon \ll 1$, it is justified to take only the $j = \pm 1$ components of the grating~\cite{shallow_grating_review}; this is equivalent to setting
\begin{equation}\label{eq:cos_grating}
\epsilon_m(x) = 2 \epsilon_m^{(1)} \cos\left(\frac{2 \pi x}{d}\right).
\end{equation}
Substituting Eq.~(\ref{eq:cos_grating}) in Eq.~(\ref{eq:deriv2}), neglecting all the fast-oscillating terms and separating into two sets of equations gives % UNIFY EQS?????? what about star??
% \begin{eqnarray}\label{eq:CMT_eqs1}
%   2 i k_{c,x} E_x^+(x,t) &=& - 2 i k_{c,x} \frac{n_0}{c} E^+_t + \frac{\epsilon_m^{(1)}}{c^2} e^{i \frac{2 \pi}{d} x - 2 i k_{x,c} x} \left[2(m_t - i \omega_0 m) E^-_t - \omega_0 (\omega_0 m + 2 i m_t) E^-\right], \nonumber \\
% - 2 i k_{c,x} E_x^-(x,t) &=& - 2 i k_{c,x} \frac{n_0}{c} E^-_t + \frac{{\epsilon_m^{(1)}}^*}{c^2} e^{- i \frac{2 \pi}{d} x + 2 i k_{x,c} x} \left[2(m_t - i \omega_0 m)E^+_t - \omega_0 (\omega_0 m + 2 i m_t) E^+\right]. \nonumber
% \end{eqnarray}
% or
\begin{eqnarray}
2 i k_0 E_x^+(x,t) + 2 \left(i k_0 \frac{n_0}{c} - (m_t - i \omega_0 m) \frac{\Delta \epsilon \epsilon_m^{(1)}}{c^2} e^{- 2 i \delta k x} \right) E^+_t &=& \Delta \epsilon \frac{\epsilon_m^{(1)}}{c^2} e^{- 2 i \delta k x} \omega_0 (\omega_0 m + 2 i m_t) E^-, \nonumber
\\ \label{eq:CMT_eqs1} \\
- 2 i k_0 E_x^-(x,t) + 2 \left(i k_0 \frac{n_0}{c} - (m_t - i \omega_0 m) \frac{\Delta \epsilon {\epsilon_m^{(1)}}^*}{c^2} e^{2 i \delta k x} \right) E^-_t &=& \Delta \epsilon \frac{{\epsilon_m^{(1)}}^*}{c^2} e^{2 i \delta k x} \omega_0 (\omega_0 m + 2 i m_t) E^+. \nonumber \\ \label{eq:CMT_eqs2}
\end{eqnarray}
For modulations longer than the pulse period, the dominant term on the Right-Hand-Side is the first. Furthermore, for a weak modulation, one can neglect the correction to the pulse velocity, i.e., the second term inside the parentheses on the Left-Hand-Side. These assumptions lead to the following coupled equations
\begin{eqnarray}
E_x^+(x,t) &+& \frac{n_0}{c} E^+_t = i \kappa e^{- 2 i \delta k x} m(t) E^-, \label{eq:Ep} \\
E_x^-(x,t) &-& \frac{n_0}{c} E^-_t = - i \kappa^* e^{2 i \delta k x} m(t) E^+, \label{eq:Em}
\end{eqnarray}
where
\begin{equation}
\kappa \equiv \Delta \epsilon \frac{\epsilon_m^{(1)}}{2 c n_0} \omega_0, %  = \frac{\pi \Delta \epsilon \epsilon_m^{(1)}}{n_c \lambda??????? \cos \theta}
\end{equation}
is a complex coupling coefficient. %  which encodes information on the strength of the modulation. 
The final form of the coupled equations~(\ref{eq:Ep})-(\ref{eq:Em}) is the same as in~\cite{shallow_grating_review} except for the nature of the modulation. Indeed, whereas in~\cite{shallow_grating_review} the modulation is induced by the traveling pulses and thus is spatio-temporally-localized, in the current context, the modulation occurs everywhere at the same time, but for a brief moment.

% Obviously, the efficiency is highest for $\delta k = 0$.

% In order to derive that solution, note that the characteristics of Eqs.~(\ref{eq:f_uniform})-(\ref{eq:b_uniform}) are solutions of
% \begin{equation}\label{eq:characteristics_full_f}
% dx = \pm \frac{v dt}{1 + ???? m(t)}, \quad \quad v = \frac{c}{n_0}.
% \end{equation}
% We integrate and define
% \begin{equation}\label{eq:moving_frames_transform}
% x^{(f,b)} \equiv x \mp v \int_{-\infty}^{t} \frac{dt'}{1 + m(t')}. \nonumber
% \end{equation}
% This is equivalent to transforming into frames moving with each of the pulses. Then, Eqs.~(\ref{})-(\ref{}) reduce to
In order to solve equations~(\ref{eq:Ep})-(\ref{eq:Em}), we follow the procedure outlined in~\cite{Sivan-Pendry-letter,Sivan-Pendry-article}. Specifically, we first transform each equation into a frame moving with each of the pulses, namely, we define $x^{(f,b)} \equiv x \mp v t$. % ,
% so that
% \begin{eqnarray}
% f_{t}(x^f(x,t),t) = - v f_{x^{(f)}} + f_{t}, \quad \quad b_{t}(x^{(b)}(x,t),t) = v b_{x^{(b)}} + b_{t}. \nonumber
% \end{eqnarray}
% substitute into Eqs.~(\ref{eq:Ep})-(\ref{eq:Em}) and get
% \begin{eqnarray}
% E^+_t(x^{(f)},t) = i v \kappa e^{- 2 i \delta k x} m(t) E^-(x^{(b)},t), \label{eq:Ept} \\
% E^-_t(x^{(b)},t) = i v \kappa^* e^{2 i \delta k x} m(t) E^+(x^{(f)},t). \label{eq:Emt}
% \end{eqnarray}
We then assume that the coupling is weak and neglect the coupling term on the RHS of Eq.~(\ref{eq:Ep}). In this case, the solution of Eq.~(\ref{eq:Ep}) is simply
\begin{equation}\label{eq:}
E^+(x^{(f)},t) = E^+(x - vt,0) \equiv E_0^+(t),
\end{equation}
where $E_0^+(t)$ is the incident pulse profile. Substituting in Eq.~(\ref{eq:Em}) % now gives
% \begin{equation}\label{eq:}
% E^-_t(x^{(b)},t) % = i v \kappa e^{- 2 i \delta k (x^{(b)} - v t)} m(t) E^+(x^{(f)},0)
% = i v \kappa e^{2 i \delta k (x^{(b)} - v t)} m(t) E^+(x^{(b)} - 2 v t). \nonumber
% \end{equation}
then gives
\begin{equation}\label{eq:E-_sol}
E^-(x^{(b)},t) = i v \kappa^* e^{2 i \delta k x^{(b)}} \int_{-\infty}^t e^{- 2 i \delta k v t'} m(t' - t_0) E^+(x^{(b)} - 2 v t') dt'.
\end{equation}
Eq.~(\ref{eq:E-_sol}) shows that the backward wave is % essentially 
given by the convolution of the forward wave with the modulation. Thus, as noted in~\cite{Sivan-Pendry-letter,Sivan-Pendry-article}, this reversal scheme yields a {\em wave-front reversal} rather than a complete time-reversal (which requires also the {\em conjugation} of the envelope).

The convolution integral can be solved exactly for a unit amplitude Gaussian pulse and a Gaussian modulation
\begin{equation}% \label{eq:f_inc_gaussian}
E^+_0\left(\frac{x}{v T_p}\right) = e^{- \frac{x^2}{v^2 T_p^2}}, \nonumber \quad \quad
% \end{equation}
% 
% \begin{equation}\label{eq:gaussian_mod}
m\left(t - t_0\right) = e^{- \frac{(t - t_0)^2}{T_{mod}^2}}. \nonumber
\end{equation}
In this case, following~\cite{Sivan-Pendry-article}, it can be shown that for $t_0 = 0$ and $\delta k = 0$, the wave-front of the reversed component is given by
\begin{eqnarray}\label{eq:E-sol}
|E^-(x + v t)| &=& \sqrt{\pi} \frac{\Delta n}{n_0} \omega_0 \frac{T_{mod} T_p}{\sqrt{T_p^2 + 4 T_{mod}^2}} E^+_0\left(\frac{x + v t}{v \sqrt{T_p^2 + 4 T_{mod}^2}}\right).
\end{eqnarray}
where $\Delta n \cong \frac{|\epsilon_m^{(1)}| \Delta \epsilon}{2 n_0}$ is the depth of the refractive index modulations. % VERIFY EXPRESSIONS!!! %  and
% \begin{equation}\label{eq:Teff}
% \frac{1}{T_{eff}^2} = \frac{1}{T_{mod}^2} + \frac{4}{T_p^2}.
% \end{equation}
% ????????? More generally, arbitrary asymmetric pulses will also undergo some distortion due to that convolution. These effects can be minimized in the non-adiabatic limit $T_{mod} \ll T_p$~\cite{Sivan-Pendry-letter,Sivan-Pendry-article}, where we get
% \begin{eqnarray}\label{eq:E-_sol_non_ad}
% E^-(x^{(b)},t) &=& i \sqrt{\pi} T_{mod} v \kappa E^+_0\left(\frac{{x^{(b)}} - 2 v t}{v T_p}\right). \nonumber
% \end{eqnarray}
% this yields an overall reversal amplitude of
% \begin{eqnarray}\label{eq:E-_sol_non_ad}
% max|E^-| &=& \sqrt{\pi} \omega_0 T_{mod} |\epsilon_m^{(1)}| \frac{\Delta \epsilon}{2 n_0^2} e^{- \frac{\Delta^2}{4} T_{mod}^2}, \nonumber
% \end{eqnarray}
% or
% , the reversal efficiency is
% \begin{eqnarray}\label{eq:E-_sol_non_ad2}
% max|E^-| &=& \sqrt{\pi} \omega_0 T_{mod} \frac{\Delta n}{n_0} e^{- \left(\frac{\Delta T_{mod}}{2}\right)^2}, \nonumber
% \end{eqnarray}
%
% \begin{eqnarray}\label{eq:dn}
% n = \sqrt{\epsilon} = n_0 \sqrt{1 + \epsilon_m m \frac{\Delta \epsilon}{n_0^2}} \cong n_0 \left(1 + \epsilon_m m \frac{\Delta \epsilon}{2 n_0^2}\right) = n_0 + \epsilon_m m \frac{\Delta \epsilon}{2 n_0}.
% \end{eqnarray} 
Note that at the time at which the modulation is maximal (i.e., at $t = t_0 = 0$), the width of the gap opened by the modulation is given approximately by $% \frac{4 c}{d} \sin^{-1}\left(\frac{\Delta \epsilon}{4 n_0^2}\right) \cong 
\frac{2 c}{d} \frac{\Delta n}{n_0}$~\cite{Yeh-book}. Thus, the reversal efficiency is proportional to the width of the gap opened by the modulation, in agreement with our interpretation of the scheme (see Section~\ref{sec:principles}).

By comparing the reversal efficiency~(\ref{eq:E-sol}) to that obtained in a 1D zero-gap PhC~\cite[Ch. VI]{Sivan-Pendry-article}, it is seen that the reversal in the HSM is about one to two orders of magnitude more efficient (in terms of intensity) than in a ZGSM. Furthermore, in comparison to the early suggestion of time-reversal via non-adiabatic 4WM~\cite{Miller_OL,Tsang_Psaltis2}, the reversal efficiency in our case is almost an order of magnitude more efficient. Thus, by splitting the time-reversal into two consecutive steps (envelope reversal via a SM and phase conjugation), one gains flexibility, simplicity (e.g., by allowing to work with a single pump pulse) and efficiency. In fact, with relatively small intensity, the phase conjugation step can be used also for amplifying the reversed signal, thus allowing for $100\%$ or more time-reversal of ultrashort optical pulses. Finally, the SM-based techniques can be implemented with a wider variety of modulation techniques, including linear modulators~\cite{Fan-reversal,Longhi-reversal,spin-wave-reversal-chumak}.

\section{Implementation}\label{sec:implementation}
As discussed in detail in~\cite{Sivan-Pendry-article}, the pulse duration dictates the required modulation technique. For pulses longer than a few picoseconds, the non-adiabatic modulation can be performed electronically, e.g., with linear electro-optical modulators. Thus, a time-reversal mirror can be built from a standard material such as $LiNbO_3$ which is spatially-modulated in a periodic manner. % Accordingly, the energies required for the modulation are significantly lower than those required for the pumps in wave-mixing based schemes. BE CAREFUL BC IF CAN DO PC WITH 3WM, THEN ENERGIES GO DOWN!
A $100\%$ reversal efficiency can be easily obtained using index modulations on the scale of $10^{-3}$~\cite{Sivan-Pendry-letter,Sivan-Pendry-article}. % , accessible in a a variety of materials.

For shorter pulses, the required modulations can be performed via a nonlinear process such as Cross-Phase Modulation or carrier injection in a pump-probe configuration. One way to ensure that the pump induces an index modulation in the $x$ direction only is to confine the probe (signal) into a thin waveguide and send a much wider and much shorter intensity-modulated pump at right angles to the waveguide (see e.g.,~\cite{Miller_OL,Tsang_Psaltis2}). 
% The pump (central) wavelength $\lambda_{pump}$ should be chosen such that the pump opens a gap at the (central) probe wavelength $\lambda_{pr}$, i.e.,
% \begin{equation}
% \lambda_{pr} = 2 n_0(\lambda_{pr}) \frac{\lambda_{pump}}{n_0(\lambda_{pump})}.
% \end{equation}
% wavelengths in vacuum.
The intensity modulation can be performed by sending a single pump pulse through a waveguide array or by interference of two pump beams. An alternative is to use a structure with a periodically-varying Kerr coefficient~\cite{Sivan_PD_06,Sivan_PRL_06}, preferably, with a changing sign (see~\cite{Boris_review_periodic_kerr} for a review).

% {\bf for $\chi^{(3)}$ applications, XPM and PC may occur simultaneously. Need to think what would a periodic modulation do to the PC effect / avoid a simultaneous nonlinearity... . In both techniques, one needs a non-adiabatic pump (or pump pulse trains).

% How would Miller work for a beam in $y$?? probably as good as our technique. }

% Finally, the medium must possess a strong cubic nonlinearity at $\lambda_{pr}$. % Some possible media are organic compounds with a high Kerr coefficient (see e.g.,~\cite{}), chalcogenide glasses~\cite{} or semiconductors.

% $n_{LiNbO_3}(\lambda = 1550nm) = 2.14$.

% chalcogenide $n_2 \sim 10^{-13} cm^2/W$, but also chalcogenide $n_2 \sim - 10^{-11} cm^2/W$!!?

% practically, i need to understand whether i will have simultaneous 4WM and XPM.


\begin{thebibliography}{99}

\bibitem{spin-wave-reversal-chumak}
A.V. Chumak, V.S. Tiberkevich, A.D. Karenowska, A.A. Serga, J.F. Gregg, A.N.
  Slavin, and B.~Hillebrands.
\newblock ``All-linear time-reversal by a dynamic artificial crystal,''
\newblock Nat. Comm. {\bf 1,} 141 (2010).

\bibitem{Fink_review}
M.~Fink.
\newblock ``Time-reversed acoustics,''
\newblock Scientific American {\bf 91}, (November 1999). 
M.~Fink.
\newblock ``Acoustic time-reversal mirrors: Imaging of complex media with
  acoustic and seismic waves,''
\newblock Topics in Applied Physics {\bf 84,} 17 (2002).

\bibitem{Mosk-Katz}
J.~Aullbach, B. Gjonaj, P.M. Johnson, A.M. Mosk and A. Lagendijk.  
\newblock ``Control of light transmission through opaque scattering media in space and time,''
\newblock Phys. Rev. Lett. {\bf 106,} 103901 (2011).
O. Katz, Y. Bromberg, E. Small and Y. Silberberg. 
\newblock ``Focusing and compression of ultrashort pulses through scattering media,''
\newblock Nat. Phot. {\bf ,} accepted. http://arxiv.org/abs/1012.0413. %  (2011).

\bibitem{Pepper-book}
D.M. Pepper.
\newblock Laser handbook, Vol. 4, p. 333.
\newblock North-Holland Physics, Amsterdam (1988).

\bibitem{tr_super_lens_pendry}
J.B. Pendry.
\newblock ``Time-reversal and negative refraction,''
\newblock Science {\bf 71,} 322 (2008).

\bibitem{Stockman_reversal}
X.~Li and M.I. Stockman.
\newblock ``Highly efficient spatio-temporal coherent control in nanoplasmonics
  on a nanometer-femtosecond scale by time-reversal,''
\newblock Phys. Rev. B {\bf 77,} 195109 (2008).

\bibitem{yaqoob_Nat_Phot}
Z.~Yaqoob, D.~Psaltis, M.S. Feld, and C.~Yang.
\newblock ``Optical phase conjugation for turbidity suppression in biological
  samples,''
\newblock Nat. Phot. {\bf 2,} 110 (2008).

\bibitem{magnonic_crystals_transmission_lines}
Y.~Kajiwara, K.~Harii, S.~Takahashi, J.~Ohe, K.~Uchida, M.~Mizuguchi,
  H.~Umezawa, H.~Kawai, K.~Ando, K.~Takanashi, S.~Maekawa, and E.~Saitoh.
\newblock ``Transmission of electrical signals by spin-wave interconversion in a
  magnetic insulator,''
\newblock Nature {\bf 464,} 262 (2010).

\bibitem{Cucchietti}
F.M. Cucchietti.
\newblock ``Time-reversal in an optical lattice,''
\newblock J. Opt. Soc. Am. B {\bf 27,} A30 (2010).

\bibitem{Miller_OL}
D.A.B Miller
\newblock ``Time reversal of optical pulses by four-wave mixing,''
\newblock Opt. Lett. {\bf 5,} 300 (1980).

\bibitem{Weiner_holography}
A.M. Weiner, D.E. Leaird, D.H. Reitze, and E.G. Paek.
\newblock ``Femtosecond spectral holography,''
\newblock IEEE J. Qu. Electron. {\bf 28,} 2251 (1992).

\bibitem{Marom_Fainman}
D.~Marom, D.~Panasenko, R.~Rokitski, P.-C. Sun, and Y.~Fainman.
\newblock ``Time-reversal of ultrafast waveforms by wave mixing of spectrally
  decomposed waves,''
\newblock Opt. Lett. {\bf 25,} 132 (2000).

% \bibitem{Marom_Fainman_reply}
% D.~Marom, D.~Panasenko, R.~Rokitski, P.-C. Sun, and Y.~Fainman.
% \newblock ``Reply to ``comment on {T}ime-reversal of ultrafast waveforms by wave mixing of spectrally decomposed waves'',''
% \newblock Opt. Lett. {\bf 25,} 1209 (2000).

\bibitem{Kuzucu_Gaeta}
O. Kuzucu and Y. Okawachi and R. Salem and M.A. Foster and A.C. Turner-Foster and M. Lipson and A.L. Gaeta.
\newblock ``Spectral phase conjugation via temporal imaging,''
\newblock Opt. Exp. \textbf{17}, 20605 (2009).

\bibitem{Tsang_Psaltis2}
M.~Tsang and D.~Psaltis.
\newblock ``Spectral phase conjugation with cross-phase modulation compensation,''
\newblock Opt. Exp. {\bf 12,} 2207 (2004). 

% \bibitem{Tsang_Psaltis3}
% M.~Tsang and D.~Psaltis.
% \newblock ``Spectral phase conjugation by quasi-phase-matched three-wave mixing,''
% \newblock Opt. Commun. {\bf 242,} 659 (2004).

% \bibitem{Tsang_4}
% M.~Tsang.
% \newblock ``Spectral phase conjugation via extended phase-matched,''
% \newblock J. Opt. Soc. Am. B {\bf 23,} 861 (2006).

\bibitem{Fan-reversal}
M.F. Yanik and S.~Fan.
\newblock ``Time-reversal of light with linear optics and modulators,''
\newblock Phys. Rev. Lett. {\bf 93,} 173903 (2004).

\bibitem{Longhi-reversal}
S.~Longhi.
\newblock ``Stopping and time-reversal of light in dynamic photonic structures
  via {B}loch oscillations,''
\newblock Phys. Rev. E {\bf 75,} 026606 (2007).

\bibitem{Matsko}
A.B. Matsko, Y.V. Rostovtsev, O. Kocharovskaya, A.S. Zibrov and M.O. Scully.
\newblock ``Nonadiabatic approach to quantum optical information storage,''
\newblock Phys. Rev. A {\bf 64,} 043809 (2001).

\bibitem{Sivan-Pendry-letter}
Y.~Sivan and J.B. Pendry.
\newblock ``Time-reversal in dynamically-tuned zero-gap periodic systems,''
\newblock Phys. Rev. Lett., {\bf 106,} 193902 (2011).
  [physics.optics].

\bibitem{Sivan-Pendry-article}
Y.~Sivan and J.B. Pendry.
\newblock ``Theory of wave-front reversal of short pulses in dynamically-tuned
  zero-gap periodic systems,'' submitted; available on ArXiv. %  to Phys. Rev. A.
  
\bibitem{shallow_grating_review}
C.M. de~Sterke and J.E. Sipe.
\newblock in Prog. in Opt. XXXIV, p. 203.
\newblock North-Holland, Amsterdam, 1994.

\bibitem{deep_gratings_PRE_96}
C.M. de~Sterke, D.G. Salinas, and J.E. Sipe.
\newblock ``Coupled-mode theory for light propagation through deep nonlinear
  gratings,''
\newblock Phys. Rev. E {\bf 54,} 1969 (1996).

\bibitem{Yeh-book}
P.~Yeh.
\newblock {\em Optical Waves in Layered Media}.
\newblock Wiley-Interscience, 2nd edition (2005).

\bibitem{Sivan_PD_06}
G.~Fibich, Y.~Sivan, and M.I. Weinstein.
\newblock ``Bound states of nonlinear Schr\"odinger equations with a periodic
  nonlinear microstructure,''
\newblock Physica D {\bf 217,} 31 (2006).

\bibitem{Sivan_PRL_06}
Y.~Sivan, G.~Fibich, and M.I. Weinstein.
\newblock ``Waves in nonlinear lattices: ultrashort optical pulses and
  bose-einstein condensates,''
\newblock Phys. Rev. Lett. {\bf 97,} 193902 (2006).

\bibitem{Boris_review_periodic_kerr}
B.A.~Malomed Y.V.~Kartashov and L.~Torner.
\newblock ``Solitons in nonlinear lattices,''
\newblock Rev. Mod. Phys. {\bf 83,} 247 (2011).

\end{thebibliography}
\end{document}